\documentclass[12pt,english,twoside]{article}

\usepackage[T1]{fontenc}
\usepackage[utf8]{inputenc}
\usepackage[letterpaper, margin=2cm]{geometry}
\usepackage{color}
\usepackage{graphicx}
\usepackage{subfigure}
\usepackage{siunitx}
\graphicspath{{graphics_fold/}}
\usepackage{times}
\usepackage{etoolbox}
\usepackage{amsmath} 
\usepackage{amssymb}  
\usepackage{fancyhdr}
\usepackage{psfrag}
\usepackage{enumitem}
\usepackage{indentfirst}       
\usepackage{titlesec}          
\usepackage{hyperref}
\usepackage{babel}
\usepackage{listings}
\usepackage{booktabs}

\usepackage{pgfplots}%
	\usepgfplotslibrary{units}
	\usepgfplotslibrary{groupplots}
	\usetikzlibrary{positioning}
	\pgfplotsset{compat=1.9}
\pgfplotsset{
 unit code/.code 2 args=
   \begingroup
   \protected@edef\x{\endgroup\si{#2}}\x
}
\usepgfplotslibrary{colorbrewer} 
\usetikzlibrary{pgfplots.colorbrewer}
\usepgfplotslibrary{colormaps} 
\usetikzlibrary{pgfplots.colormaps}
\usepackage{pgfplots,tikz}

\addto\captionsenglish{}

\usepackage[colorinlistoftodos]{todonotes} 

\setlist{nolistsep}

\makeatletter
\renewcommand*{\@biblabel}[1]{\hfill#1.}
\makeatother

\fancypagestyle{firstpage}{
\fancyhead{}
\fancyfoot{}
\fancyfoot[LE,RO]{\small \thepage}

}

\makeatletter
\def\maketitle{
  \thispagestyle{firstpage}
  {
   \smallskip
   \fontsize{18}{20}\selectfont\sffamily{}  \noindent \MakeUppercase{\textbf{\@title}}

   \bigskip
   \fontsize{14}{20}\selectfont\rmfamily{} \noindent \@author
  }
}
\makeatother

\titleformat{\section}
  {\fontsize{14}{14}\normalfont\sffamily\bfseries}
  {\thesection.}{1em}{}                

\titleformat{\subsection}
  {\normalfont\sffamily\bfseries}
  {\thesubsection}{1em}{}

\titleformat{\subsubsection}
  {\normalfont\sffamily\small}
  {\textit{\thesubsubsection}}{1em}\textit{{}}

\title{Feature-based representation for violin bridge admittances}

\author{Raffaele Malvermi, Sebastian Gonzalez, Martino Quintavalla, Fabio Antonacci\\and Augusto Sarti\\
{\small \textit{Departement of Electronics, Information and Bioengineering, Politecnico di Milano, Milano, Italy\\
e-mail: raffaele.malvermi@polimi.it}}\\
Jesús Alejandro Torres\\
{\small \textit{Laboratory of Acoustics, Violin Making School of Mexico, Sub-directorate General for Education and Artistic Research, National Institute of Fine Arts and Literature, Hidalgo \#20 Centro, Querétaro, Mexico CP 76000}}\\
Roberto Corradi\\
{\small \textit{Departement of Mechanical Engineering, Politecnico di Milano, Milano, Italy}}}

\begin{document}

\maketitle
\renewcommand{\abstractname}{\vspace{-\baselineskip}} 

\begin{abstract}	\noindent
Frequency Response Functions (FRFs) are one of the cornerstones of musical acoustic experimental research.
They describe the way in which musical instruments vibrate in a wide range of frequencies and are used to predict and understand the acoustic differences between them.
In the specific case of stringed musical instruments such as violins, FRFs evaluated at the bridge are known to capture the overall body vibration.
These indicators, also called bridge admittances, are widely used in the literature for comparative analyses.
However, due to their complex structure they are rather difficult to quantitatively compare and study.
In this manuscript we present a way to quantify differences between FRFs, in particular violin bridge admittances, that separates the effects in frequency, amplitude and quality factor of the first resonance peaks characterizing the responses.
This approach allows us to define a distance between FRFs and clusterise measurements according to this distance. We use two case studies, one based on Finite Element Analysis and another exploiting measurements on real violins, to prove the effectiveness of such representation.
In particular, for simulated bridge admittances the proposed distance is able to highlight the different impact of consecutive simulation `steps' on specific vibrational properties and, for real violins, gives a first insight on similar styles of making, as well as opposite ones.
\\

\end{abstract}

\quad\rule{425pt}{0.4pt}

\section{Introduction}
\label{sec:introduction}
Comparing musical instruments is always a hard task to perform due to the wide range of perspectives involved.
Violins can be described by either trying to define their timbre and thus focusing on the sound perceived or looking at the specific models used by luthiers in their making.
These analyses are often affected by subjectivity, leading to a whole literature of blind tests to assess the variability of human judgements \cite{fritz2010perceptual,fritz2012player}.
More objective representations exploit indicators describing the vibrational properties of musical instruments \cite{corradi2015multidisciplinary}.
In the case of violins, the bridge admittance is a good estimator for structural vibration thanks to its reliability, but still too redundant to be adopted for advanced classification and clustering tasks.
Nonetheless, recent works in the context of musical acoustics still analyze these signals through visual comparisons \cite{freour2020trumpet}.

Once a set of FRFs is collected from different musical instruments, a distance metric has to be devised in order to perform objective comparisons.
Point measures based on p-norms such as Mean Square Error (MSE) and Mean Absolute Value (MAE) are widely used in signal processing applications but they work under the implicit assumption that all samples contribute in the same way to assess similarity \cite{wang2009mse}.
This is not the case in violin bridge admittances where the amplitude, the frequency and the damping factor characterizing the resonances of the system seem to be more informative.

The literature showed the benefits of evaluating such indicators to highlight dependencies between mechanical vibration, geometry, material properties and chemical treatments. In particular, it is well known that the solution of the partial differential equation, i.e.~the eigenmodes, governing structural vibration strongly depends on the boundary conditions imposed, hence on the system geometry \cite{meirovitch2001fundamentals}.
Changes in the elastic properties of the material under analysis influence the distribution of natural frequencies and some mathematical relations have been found for wooden plates between specific eigenfrequencies and Young's and shear moduli \cite{mcintyre1986woodproperties3,sprossman2017materialproperties}.
On the other hand, damping has proved to be sensitive to changes imposed on the material by means of chemical and heat treatments \cite{yano1994dampingtreatment} and, in the specific case of wood, also to its moisture content \cite{goken2018dampingmoisture}.
Finally, although the estimated resonance peak amplitude suffers low sensitivity to changes due to the intrinsic variability characterizing hammer test acquisitions, it can give hints on variations related to mode shapes.
This makes this feature meaningful when the similarity is evaluated between periodic measurements on the same system, either for health monitoring purposes or for tracking conscious design modifications \cite{mihualcicua2020frequency,corradi2017pianotracking}.

In this work, we propose the definition of a multidimensional feature space based on modal parameters where a more informative Euclidean distance can be measured to assess the similarity of FRFs in the low frequency range.
The soundness of the proposed methodology (Section~\ref{sec:methodology}) has been tested on two different case studies, based on simulated (Section~\ref{sec:experiments_titian}) and real (Section~\ref{sec:experiments_cremona}) responses, respectively.

\section{Methodology}
\label{sec:methodology}
Frequency Response Functions (FRFs) describe point-to-point relations between applied force and consequent vibration in a mechanical structure.
Given the aforementioned physical quantities measured at two points $i$ and $j$, the resulting FRF
\begin{equation}\label{eq:frf}
    H_{ij}(\omega) = \frac{X_j(\omega)}{F_i(\omega)}    
\end{equation}
gives local information about the system response as a function of frequency $\omega$.
When $F_i(\omega)$ and $X_j(\omega)$ are the spectra associated to a force applied at point $i$ and a velocity acquired at point $j$, $H_{ij}(\omega)$ is called \textit{admittance} or \textit{mobility}.
In experimental applications, such as those involving violin bridge admittances, FRFs are estimated exploiting a set of redundant measurements by means of the so-called H1, H2 and Hv estimators.
These estimators are chosen depending on the assumptions concerning how to model the presence of noise in the signals \cite{schwarz1999emabasics}.

Recalling that violin bridge admittances are characterized by low modal density in the low frequency range, let us consider a mathematical description based on modal superposition \cite{meirovitch2001fundamentals}, namely
\begin{equation}\label{eq:modal_frf}
    H_{ij}(\omega) \sim \sum_{r=1}^N \frac{\Phi_{ri}\Phi_{rj}}{\omega_r^2 + 2j\omega\xi_r\omega_r - \omega^2},
\end{equation}
where $N$ is the number of modes considered, $\Phi_{ri}$ and $\Phi_{rj}$ correspond to the mode shape of mode $r$ evaluated at the two points observed, $\omega_r$ is the natural frequency and $\xi_r$ is the damping ratio associated to mode $r$, respectively.
It is worth noticing that when $\omega$ matches $\omega_r$,
\begin{equation}\label{eq:modal_magnitude}
    |H_{ij}(\omega_r)| \sim \frac{\Phi_{ri}\Phi_{rj}}{2\xi_r\omega_r^2},
\end{equation}
thus making Eq.~\eqref{eq:modal_frf} completely recovered from the knowledge of the mode frequencies, amplitudes and corresponding damping ratios.
This work proposes to project signals into a new feature space where the parameters just mentioned constitute the set of observables and Eq.~\eqref{eq:modal_frf} acts as a mapping function.

\begin{figure}[t]
    \centering
    \includegraphics[width=\textwidth]{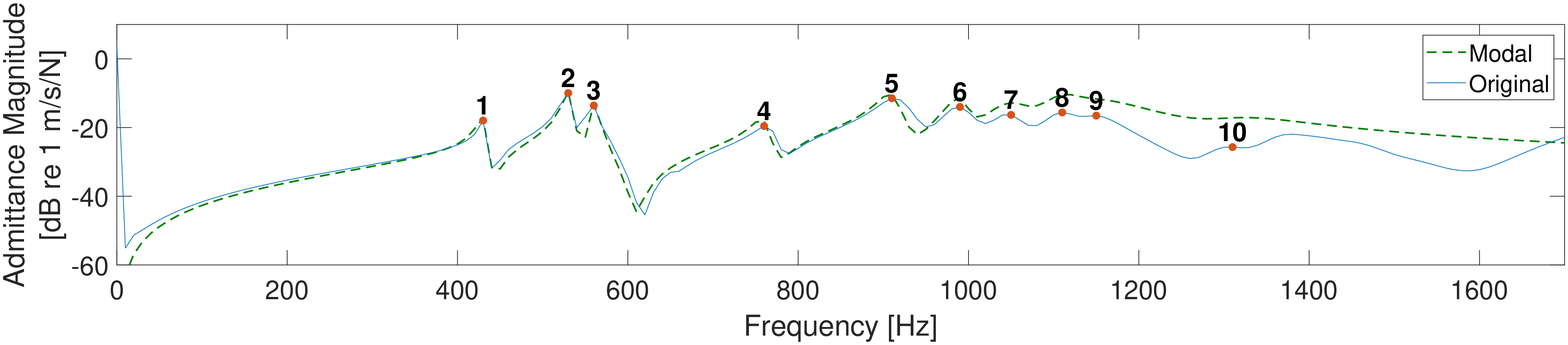}
    \caption{Example of FRF reconstruction starting from the extracted features. In blue the original signal, in green its modal approximation. The first ten peaks detected are highlighted with red dots.}
    \label{fig:test_modal}
\end{figure}

The feature extraction process is developed as follows.
Given a set of $M$ admittances (for fixed input and output positions $i$ and $j$), the first $N=10$ resonances are identified with the aid of a suitable peak finding algorithm and the resulting coordinates expressed in frequency and amplitude (in dB scale) are collected inside two matrices $\mathbf{F},\mathbf{P}\in\mathbb{R}^{M\times N}$.
Features related to damping are estimated in terms of Q Factor by means of the half-power bandwidth method \cite{chopra2019fundamentals} and stored in a third matrix $\mathbf{Q}\in\mathbb{R}^{M\times N}$. In order to better evaluate the peak bandwidth, parabolic interpolation has been performed on samples in the proximity of each peak.
Fig.~\ref{fig:test_modal} shows an example of signal reconstruction starting from the features extracted.
The approximation, represented through the green dotted line, is the evaluation of Eq.~\eqref{eq:modal_frf} based on the modal features stored in $\mathbf{F},\mathbf{P}$ and $\mathbf{Q}$.
Since Eq.~\eqref{eq:modal_frf} assumes distinct peaks and low damping, it can be noticed that the accuracy of the reconstructed signal decreases when peaks are closer and more correlated.

The resulting feature space considered has thus dimensionality $3N$.
Feature normalization \cite{jain2005zscore} has been applied over each feature vector $\mathrm{f},\mathrm{p},\mathrm{q}$, i.e. the columns of $\mathbf{F}, \mathbf{P}$ and $\mathbf{Q}$, in order to avoid bias.
Once normalized, an Euclidean distance is evaluated on the feature vectors characterizing each pair of admittances, leading to a distance matrix $\mathbf{D}\in\mathbb{R}^{M\times M}$ with elements
\begin{equation}\label{eq:distance}
    [D]_{n,m} = \|\mathrm{f}_n-\mathrm{f}_m\|_2 +
    \|\mathrm{p}_n-\mathrm{p}_m\|_2 + 
    \|\mathrm{q}_n-\mathrm{q}_m\|_2,
\end{equation}
where $n,m=1,...,M$.
Addends on the right side of Eq.~\eqref{eq:distance} can be also analysed separately to highlight variations in a specific feature subspace.

\section{Results}
\label{sec:experiments}
To show the effectiveness of the feature-based representations presented in Section~\ref{sec:methodology}, two case studies are proposed.
The first dataset is composed by bridge admittances predicted by means of Finite Element Analysis (FEA) using the Stradivarius' \textit{Titian} violin as reference \cite{torres2020titian}.
The second application analyzes similarities between experimental acquisitions taken on a set of contemporary violins.

\subsection{FEA simulations of Stradivarius' Titian}\label{sec:experiments_titian}
Variations in signature modes of violins due to changes in the soundbox geometry were analyzed in \cite{torres2020titian}.
An efficient parametric FEA model was defined in order to perform before/after comparisons concerning frequency responses and mode shapes.
One of the instruments belonging to Stradivarius' golden age, the \textit{Titian}, was used as reference for the model development.
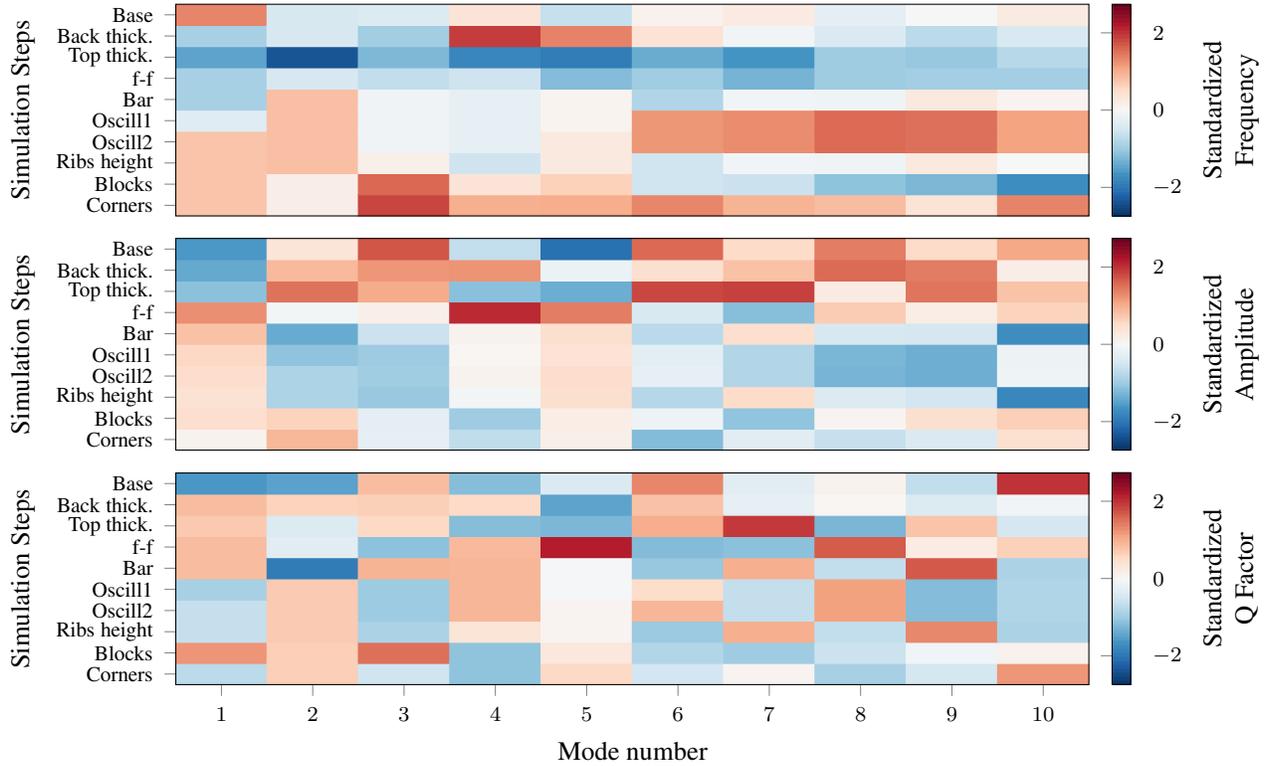
\begin{figure}[t]
    \centering
\pgfplotsset{
    colormap/RdBu,
    colormap={RdBu-reversed}{
        indices of colormap=(
            \pgfplotscolormaplastindexof{RdBu},...,0 of RdBu
            )
    },
}
\begin{tikzpicture}
    \begin{groupplot}[
            group style={group size= 1 by 3, group name=myGroup, vertical sep=0.3cm}, 
            height=0.25\textwidth, 
            width=0.78\textwidth,
            label style={font=\footnotesize},
            tick label style={font=\scriptsize},
            xtick pos=bottom,ytick pos=left,
            every tick label/.append style={font=\scriptsize},
            tick align=outside,
            xtick={1, 2, 3, 4, 5, 6, 7, 8, 9, 10},
            ytick={1, 2, 3, 4, 5, 6, 7, 8, 9, 10},
            yticklabels={Corners,Blocks,Ribs height,Oscill2,Oscill1,Bar,f-f,Top thick.,Back thick.,Base},
            colorbar style={
                yticklabel style={
                   /pgf/number format/fixed,
                },
                width=7pt,
                point meta min = -2.7394,
                point meta max = 2.7394},
            enlarge x limits={abs=0.5},
            enlarge y limits={abs=0.5},
            axis on top,
            xmin=1,
            xmax=10,
            ymin=1,
            ymax=10,
                ]
            ]
        \nextgroupplot[
            ylabel=Simulation Steps,
            xticklabels = \empty,
            xtick = \empty,
            point meta min = -2.7394,
            point meta max = 2.7394,
            colorbar,
            colorbar style={
                y label style={align=center},
                ylabel=Standardized\\Frequency,
            },
            ]
            
            \addplot [matrix plot*,point meta=explicit, mesh/cols=10,
            ] file [meta=index 3]{./anc/figure2a.txt};
            
        \nextgroupplot[
            ylabel=Simulation Steps,
            xticklabels = \empty,
            xtick = \empty,
            point meta min = -2.7394,
            point meta max = 2.7394,
            colorbar,
            colorbar style={
                y label style={align=center},
                ylabel=Standardized\\Amplitude,
            },
            ]
            \addplot [matrix plot*,point meta=explicit, mesh/cols=10,
            ] file [meta=index 3]{./anc/figure2b.txt};
            
        \nextgroupplot[
            xlabel=Mode number,
            ylabel=Simulation Steps,
            point meta min = -2.7394,
            point meta max = 2.7394,
            colorbar,
            colorbar style={
                y label style={align=center},
                ylabel=Standardized\\Q Factor,
            },
            ]
            \addplot [matrix plot*,point meta=explicit, mesh/cols=10,
            ] file [meta=index 3]{./anc/figure2c.txt};
    \end{groupplot}

\end{tikzpicture}
    \caption{Feature Matrices extracted from simulated bridge admittances: from top to bottom, natural frequency ($\mathbf{F}$), peak amplitude ($\mathbf{P}$) and Q Factor ($\mathbf{Q}$) of the first $N$ modes, respectively.}
    \label{fig:titian_genetic_codes}
\end{figure}
We have projected the bridge admittance into the proposed feature space for each of the $M=10$ simulation steps presented in the original work. Each step is characterized by changes in the thickness profile, addition of components and artificial signature modes.
Indeed, a time domain velocity response has been collected at the corner of the bridge for each simulation step, after the application of an impulsive force.
Signals duration is $\SI{100}{\milli\second}$, sampled at $F_s = \SI{10}{\kilo\hertz}$.
Admittances have been obtained as the ratio between the response and excitation spectra, following the definition in Eq.~\eqref{eq:frf}.

Fig.~\ref{fig:titian_genetic_codes} shows the resulting feature vectors, organized in the $\mathbf{F}$, $\mathbf{P}$ and $\mathbf{Q}$ matrices, respectively.
Fixing a specific row in the three subplots, all the features related to a specific step can be evaluated.
In general, feature vectors related to the first five peaks are characterized by more stable patterns, as few discontinuities occur, thus highlighting correlations between steps and modes.
One example is given by the addition of the low frequency harmonic oscillators (Oscill1 and Oscill2), whose impact is clearly visible on the first peak in terms of frequency and damping since a new artificial resonance is added inside the FRF.
Inspecting matrix $\mathbf{P}$ in the same figure, it can be noticed that the addition of the f-holes has a relevant impact on the amplitudes, in particular for peaks 1, 4 and 5.
The sensitivity of FRFs to f-holes detected in the feature vectors agrees with \cite{torres2020titian}, where a systemic analysis of these components was performed.

\begin{figure}[t!]
    \centering

\begin{tikzpicture}
    \begin{groupplot}[
            group style={group size= 3 by 1, group name=myGroup, horizontal sep=0.5cm, vertical sep=3.5cm},
            height=0.31\textwidth, 
            width=0.31\textwidth,
            label style={font=\footnotesize},
            tick label style={font=\scriptsize},
            x tick label style={rotate=90},
            xtick pos=bottom,ytick pos=left,
            every tick label/.append style={font=\scriptsize},
            tick align=outside,
            xtick={10,9,8,7,6,5,4,3,2,1},
            xticklabels={Corners,Blocks,Ribs height,Oscill2,Oscill1,Bar,f-f,Top thick.,Back thick.,Base},
            ytick={1,2,3,4,5,6,7,8,9,10},
            colorbar style={
                yticklabel style={
                   /pgf/number format/fixed,
                },
                colormap/viridis,
                width=7pt},
            enlarge x limits={abs=0.5},
            enlarge y limits={abs=0.5},
            axis on top,
            xmin=1,
            xmax=10,
            ymin=1,
            ymax=10,
            ]
        \nextgroupplot[
            title=Frequency Distance, 
            xlabel=Simulation Steps,
            ylabel=Simulations Steps,
            yticklabels={Corners,Blocks,Ribs height,Oscill2,Oscill1,Bar,f-f,Top thick.,Back thick.,Base},
            point meta max = 61.4590,
        ]
            
            \addplot [matrix plot*,point meta=explicit, mesh/cols=10,colormap/viridis,
            ] file [meta=index 3]{./anc/figure3a.txt};
            
        \nextgroupplot[
            title=Amplitude Distance, 
            xlabel=Simulation Steps,
            yticklabels=\empty,
            ytick=\empty,
            point meta min = 0,
            point meta max = 61.4590,
            ]
            
            \addplot [matrix plot*,point meta=explicit, mesh/cols=10,colormap/viridis,
            ] file [meta=index 3]{./anc/figure3b.txt};
            
        \nextgroupplot[
            title=Q Factor Distance, 
            xlabel=Simulation Steps,
            yticklabels=\empty,
            ytick=\empty,
            point meta min = 0,
            point meta max = 61.4590,
            colorbar]
            
            \addplot [matrix plot*,point meta=explicit, mesh/cols=10,colormap/viridis,
            ] file [meta=index 3]{./anc/figure3c.txt};
          
    \end{groupplot}
    \node[below = 2.2cm of myGroup c1r1.south] {(a)};
    \node[below = 2.2cm of myGroup c2r1.south] {(b)};
    \node[below = 2.2cm of myGroup c3r1.south] {(c)};
    
\end{tikzpicture}
    \caption{Distance Matrices computed starting from feature vectors depicted in Fig.~\ref{fig:titian_genetic_codes}. Euclidean distances are assessed over modal frequency (a), amplitude (b) and Q factor (c), separately.}
    \label{fig:titian_distances}
\end{figure}
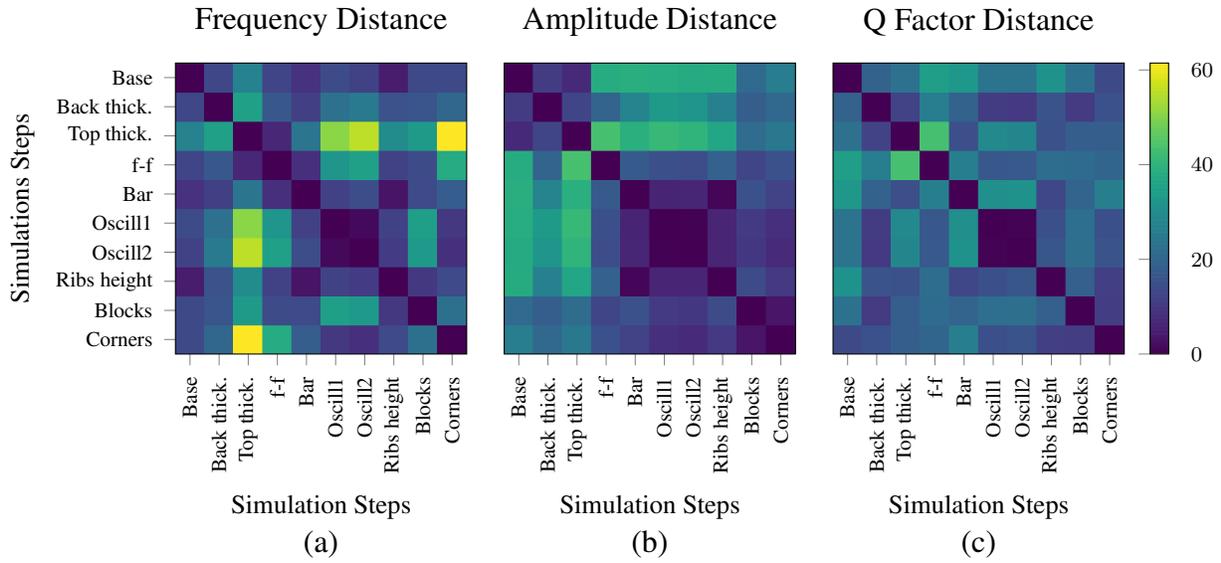

\begin{figure}[b!]
    \centering

\begin{tikzpicture}
    \begin{groupplot}[
            group style={group size= 2 by 1, group name=myGroup, horizontal sep=2cm, vertical sep=3.5cm},
            height=0.3\textwidth, 
            width=0.3\textwidth,
            label style={font=\footnotesize},
            tick label style={font=\scriptsize},
            x tick label style={rotate=90},            xtick pos=bottom,ytick pos=left,
            every tick label/.append style={font=\scriptsize},
            tick align=outside,
            xtick={10,9,8,7,6,5,4,3,2,1},
            xticklabels={Corners,Blocks,Ribs height,Oscill2,Oscill1,Bar,f-f,Top thick.,Back thick.,Base},
            ytick={1,2,3,4,5,6,7,8,9,10},
            colorbar,
            colorbar style={
                yticklabel style={
                   /pgf/number format/fixed,
                },
                colormap/viridis,
                width=7pt},
            enlarge x limits={abs=0.5},
            enlarge y limits={abs=0.5},
            axis on top,
            xmin=1,
            xmax=10,
            ymin=1,
            ymax=10,
            ]
        \nextgroupplot[
            title=Feature Distance, 
            xlabel=Simulation Steps,
            ylabel=Simulations Steps,
            yticklabels={Corners,Blocks,Ribs height,Oscill2,Oscill1,Bar,f-f,Top thick.,Back thick.,Base},
            point meta min = 0,
        ]
            
            \addplot [matrix plot*,point meta=explicit, mesh/cols=10,colormap/viridis,
            ] file [meta=index 3]{./anc/figure4a.txt};
            
        \nextgroupplot[
            title=Mean Square Error, 
            xlabel=Simulation Steps,
            yticklabels=\empty,
            ytick=\empty,
            point meta min = 0,
            ]
            
            \addplot [matrix plot*,point meta=explicit, mesh/cols=10,colormap/viridis,
            ] file [meta=index 3]{./anc/figure4b.txt};
            
    \end{groupplot}
    \node[below = 2.2cm of myGroup c1r1.south] {(a)};
    \node[below = 2.2cm of myGroup c2r1.south] {(b)};
\end{tikzpicture}
    \caption{Comparison between proposed distance metric (a) and Mean Square Error (b).}
    \label{fig:titian_comparison}
\end{figure}
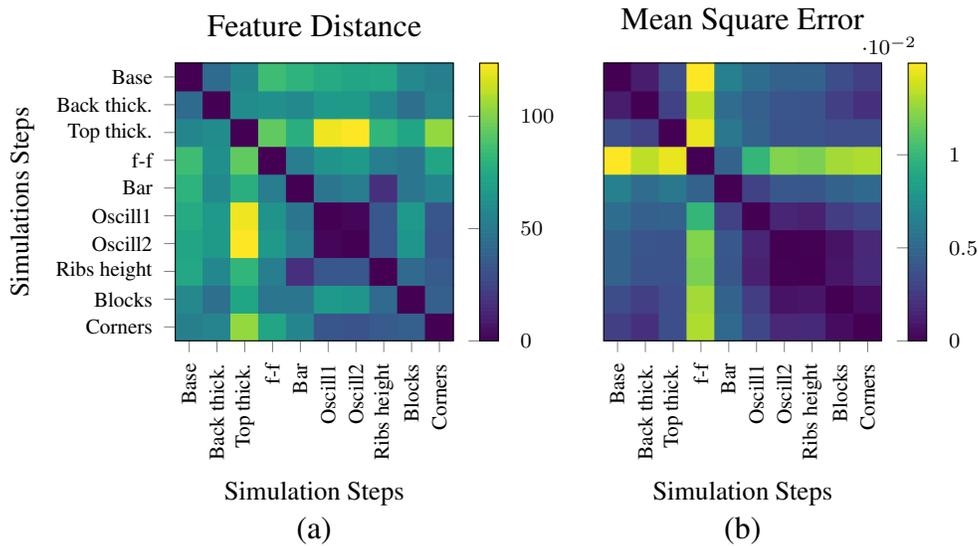

From feature vectors, distance matrices can be computed to assess either the similarity between model changes or the relevance of a specific soundbox configuration within the chain.
Fig.~\ref{fig:titian_distances} depicts the distance evaluated considering the three subspaces separately.
Resulting matrices are symmetric, due to the inherent symmetry of the euclidean distance.
Hence, it is possible to highlight clusters along the diagonal of each matrix, where the distance between consecutive steps is lower.
Inspecting the three matrices, it can be observed that the tuning of the top plate has a relevant impact on the frequency distribution, not only with respect to the previous step but also considering the whole simulation chain, while the addition of the f-holes is the most important step regarding amplitude and damping.
Moreover, distances in Fig.~\ref{fig:titian_distances}a range within a wider scale with respect to those related to amplitude and damping, thus making features based on frequency more informative.

Results depicted in Fig.~\ref{fig:titian_distances} can be combined using Eq.~\eqref{eq:distance}, leading to the matrix in Fig.~\ref{fig:titian_comparison}a.
In this representation, it can be observed that visible clusters correspond to consecutive simulation steps, where each new step has a minor or negligible impact on all the vibrometric features extracted.
In particular, the tuning of the top plate can be considered as the most discriminative step, while the two low frequency oscillators show the same distances inside the matrix, thus suggesting that the addition of the second oscillator does not yield relevant effects on the vibrational response.
Fig.~\ref{fig:titian_comparison}b shows the Mean Square Error (MSE) between the FRFs computed between consecutive steps.
It can be noticed that the MSE distance matrix is characterized by large clusters and values ranging within a small scale.
Only the addition of the f-holes highlights a different behaviour in the corresponding admittance, mostly related to a relevant amplification of the peaks, while the rest of the signal is left unchanged.
The same observation can be confirmed by inspecting the effects of the addition of f-holes in Fig.~\ref{fig:titian_distances}, where the distance from the previous step is low concerning the frequency distribution while is large in the other two feature subspaces.

From a comparison of the two distances, it is possible to observe that the proposed one achieves better discrimination capability, as it allows to identify subtle differences between consecutive steps.

\subsection{Cremonese contemporary violins}\label{sec:experiments_cremona}

\begin{figure}[b!]
    \begin{subfigure}
        \centering
\pgfplotsset{
    colormap/YlGn,
    colormap={YlGn-reversed}{
        indices of colormap=(
            \pgfplotscolormaplastindexof{YlGn},...,0 of YlGn
            )
    },
}
\begin{tikzpicture}
    \begin{groupplot}[
            group style={group size= 1 by 1, group name=myGroup},
            height=0.29\textwidth, 
            width=0.29\textwidth,
            label style={font=\small},
            tick label style={font=\small},
            x tick label style={rotate=90},
            xtick pos=bottom,ytick pos=left,
            every tick label/.append style={font=\small},
            tick align=outside,
            xtick={0.0, 1.0, 2.0, 3.0, 4.0, 5.0, 6.0, 7.0, 8.0},
            ytick={0.0, 1.0, 2.0, 3.0, 4.0, 5.0, 6.0, 7.0, 8.0},
            colorbar,
            colorbar style={
                ylabel=Feature Distance,
                yticklabel style={
                   /pgf/number format/fixed,
                },
                width=7pt},
            enlarge x limits={abs=0.5},
            enlarge y limits={abs=0.5},
            axis on top,
            xmin=0,
            xmax=8,
            ymin=0,
            ymax=8,
            ]
        \nextgroupplot[
            xlabel=Violin,
            ylabel=Violin,
            xticklabels={V5,V3,V4,V6,Strad,V1,V8,V7,V2},
            yticklabels={V2,V7,V8,V1,Strad,V6,V4,V3,V5},
        ]
            
            \addplot [matrix plot*,point meta=explicit, mesh/cols=9,
            ] file [meta=index 3]{./anc/figure5a.txt};

    \end{groupplot}

\end{tikzpicture}
    \end{subfigure}
    \begin{subfigure}
        \centering
        \includegraphics[width=.63\textwidth]{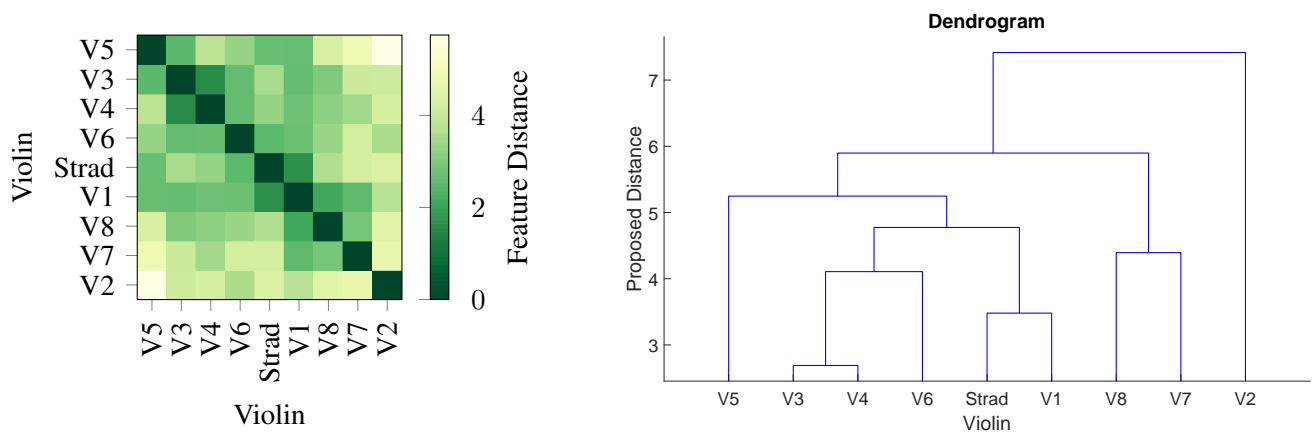}
    \end{subfigure}
    \begin{subfigure}
        \centering
        \hspace*{7.5em}
    \end{subfigure}
    \begin{subfigure}
        \centering
        \hspace*{20.1em}
    \end{subfigure} 
    \caption{Proposed distance metric (a) and corresponding dendrogram (b) computed on real bridge admittances.}
    \label{fig:cremona_tot_distance}
\end{figure}

A hammer testing campaign has been performed on 7 Cremonese contemporary violins in order to collect a new dataset of bridge admittances.
Impacts were applied on one corner of the bridge using a dynamometric hammer with light tip (086E80, by \emph{PCB Piezotronics}) and an accelerometer (352A12, by \emph{PCB Piezotronics}) was placed on the opposite side to collect the system response.
For each violin, 6 time domain acquisitions were averaged to reduce noise, following the definition of the H1 estimator \cite{schwarz1999emabasics}.
Finally, the dataset was enriched with an FRF of Stradivarius' violin and an instrument designed using its model (violin V1).
Since the sound produced by the two violins was considered similar accordingly to the owner, the addition of their bridge admittances is considered as a benchmark to verify their similarity also in terms of vibrational responses and give a proof of the soundness of the proposed representation.

Fig.~\ref{fig:cremona_tot_distance} shows the results obtained through the application of the proposed metric.
To highlight clusters, the matrix ordering adopted was devised computing the dendrogram \cite{gruvaeus1972dendrogram} of the distances obtained.
This step becomes unnecessary when the dataset is characterized by an inherent ordering, such as the case study presented in Section~\ref{sec:experiments_titian}.
It is noteworthy to observe that the Stradivarius' violin and its copy (V1) belong to the same cluster, due to the similarity between the two signals up to 1.4KHz and, in particular, their signature modes.
In the same way, other clusters characterized by low values of the distance can be detected, i.e. V3-V4 and V1-V8, respectively.

The inspection of the distance matrix can give hints also on different styles of making.
Indeed, V2 presents large distances with respect to all the other violins of the dataset, in particular when compared to V5.
By looking at the corresponding dendrogram in Fig.~\ref{fig:cremona_tot_distance}b, it can be noticed that these two violins are placed at the extremes of the tree ordering, while the old violin acts as a center of mass within the distribution of the tree leaves.

\begin{figure}[t!]
    \begin{subfigure}
        \centering
\pgfplotsset{
    colormap/YlGn,
    colormap={YlGn-reversed}{
        indices of colormap=(
            \pgfplotscolormaplastindexof{YlGn},...,0 of YlGn
            )
    },
}
\begin{tikzpicture}
    \begin{groupplot}[
            group style={group size= 1 by 1, group name=myGroup},
            height=0.29\textwidth, 
            width=0.29\textwidth,
            label style={font=\small},
            tick label style={font=\small},
            x tick label style={rotate=90},
            xtick pos=bottom,ytick pos=left,
            every tick label/.append style={font=\small},
            tick align=outside,
            xtick={0.0, 1.0, 2.0, 3.0, 4.0, 5.0, 6.0, 7.0, 8.0},
            ytick={0.0, 1.0, 2.0, 3.0, 4.0, 5.0, 6.0, 7.0, 8.0},
            colorbar,
            colorbar style={
                ylabel=Mean Square Error,
                yticklabel style={
                   /pgf/number format/fixed,
                },
                width=7pt},
            enlarge x limits={abs=0.5},
            enlarge y limits={abs=0.5},
            axis on top,
            xmin=0,
            xmax=8,
            ymin=0,
            ymax=8,
            ]
            
        \nextgroupplot[
            xlabel=Violin,
            ylabel=Violin,
            xticklabels={Strad,V5,V4,V3,V6,V2,V7,V8,V1},
            yticklabels={V1,V8,V7,V2,V6,V3,V4,V5,Strad},
        ]
            
            \addplot [matrix plot*,point meta=explicit, mesh/cols=9,
            ] file [meta=index 3]{./anc/figure6a.txt};
            
    \end{groupplot}

\end{tikzpicture}
    \end{subfigure}
    \begin{subfigure}
        \centering
        \includegraphics[width=.63\textwidth]{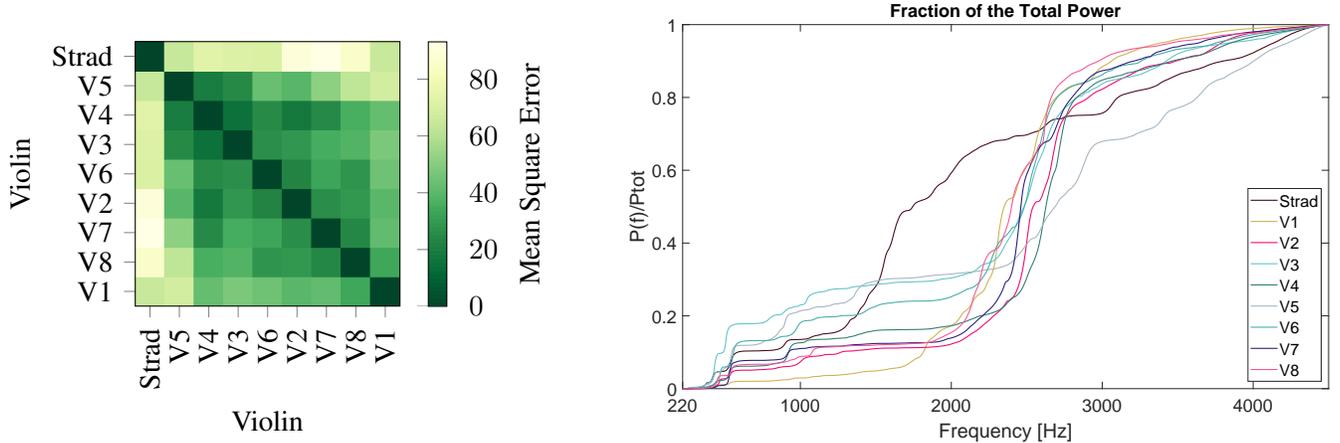}
    \end{subfigure}
    \begin{subfigure}
        \centering
        \hspace*{7.5em}
    \end{subfigure}
    \begin{subfigure}
        \centering
        \hspace*{20.1em}
    \end{subfigure} 
    \caption{MSE (a) and Fraction of the total power (b) computed on real bridge admittances.}
    \label{fig:cremona_mse}
\end{figure}

The same considerations cannot be made using the MSE as metric.
The distance matrix depicted in Fig.~\ref{fig:cremona_mse}a shows exactly the same drawbacks encountered in Section~\ref{sec:experiments_titian}, where all the signals in the dataset share similar values of the distance.
In this case, the old violin is characterized by high values of the MSE with respect to all the other musical instruments, including V1.
This undesired behavior of the MSE can be explained by the introduction of information regarding the high frequency range of signals, and in particular of the bridge hill, characterized by high modal density and variability.

An indicator of the bridge hill position in frequency can be estimated computing the power fraction of the signal, namely
\begin{equation}
    \frac{P(\omega)}{P_{tot}} = \frac{\int_{0}^{\omega} |H_{ij}(\omega)|^2 \,d\omega}{\int_{0}^{\inf} |H_{ij}(\omega)|^2 \,d\omega},
\end{equation}
where $P(\omega)$ is the power of the signal over the frequency range $[0, \omega]$ and $P_{tot}$ is the power of the whole signal.
Inspecting Fig.~\ref{fig:cremona_mse}b, it can be noticed that the bridge hill is located where the power fraction shows its maximum slope.
The old violin shows a different behavior with respect to contemporary violins, thus explaining the difference detected by MSE with respect to the other violins.
However, the weight given by MSE to the differences in the bridge hill region hides the strong similarity between the signature modes of the old violin and its copy, which is recovered by the proposed distance.

\section{Conclusions}
\label{sec:conclusions}

In this paper we have presented a new method to compare Frequency Response Functions in violins in a quantitative way.
The resulting representations can be used both for evaluating experiments and simulations.
Previously in the literature, the comparison of bridge admittances has been rather subjective and descriptive, and resembled more of a dark art than science.
By looking at certain features of the FRF signal, in particular those related to the underlying physics of the instrument, we are able to define a distance between instruments and steps in the simulation process.
This distance gave us insights as to what are the most significant processes in the simulation.
One possible application of our method is its use in the optimization of violin copies \cite{gonzalez2020a}: by defining a particular distance function one can focus on copying particular features of a violin.
This is in general easier than using point measures, such as the MSE, since we can focus on the significant features of the vibrational response.
The proposed methodology has been devised also in a real case scenario to assess the similarity between old and contemporary violins.
Results proved the soundness of such representation in assessing similar styles of making as well as opposite ones, according to judgements given by the owners.
The distance computed on features is very informative on signature modes, while it lacks knowledge on some important high frequency regions such as the bridge hill.
Considering the introduction of integral features, i.e. based on the power fraction, can extend the observation window of the proposed representation without loosing expressive power in the low frequency range.

\begingroup
    \fontsize{10pt}{10pt}\selectfont
    \bibliographystyle{ref_style}
    \bibliography{ms}

\newcommand{\noop}[1]{}
\begin{thebibliography}{10}

\bibitem{fritz2010perceptual}
Fritz, C., Woodhouse, J., Cheng, F. P.-H., Cross, I., Blackwell, A.~F. and
  Moore, B.~C.
\newblock Perceptual studies of violin body damping and vibrato, {\em The
  Journal of the Acoustical Society of America\/}, {\bf 127} (1), 513--524,
  (2010).

\bibitem{fritz2012player}
Fritz, C., Curtin, J., Poitevineau, J., Morrel-Samuels, P. and Tao, F.-C.
\newblock Player preferences among new and old violins, {\em Proceedings of the
  National Academy of Sciences\/}, {\bf 109} (3), 760--763, (2012).

\bibitem{corradi2015multidisciplinary}
Corradi, R., Liberatore, A., Miccoli, S., Antonacci, F., Canclini, A., Sarti,
  A. and Zanoni, M.
\newblock A multidisciplinary approach to the characterization of bowed string
  instruments: The musical acoustics lab in cremona, {\em ICSV22 - 22nd
  International Congress on Sound and Vibration\/}, (2015).

\bibitem{freour2020trumpet}
Fréour, V., Guillot, L., Masuda, H., Usa, S., Tominaga, E., Tohgi, Y., Vergez,
  C. and Cochelin, B.
\newblock Numerical continuation of a physical model of brass instruments:
  Application to trumpet comparisons, {\em The Journal of the Acoustical
  Society of America\/}, {\bf 148} (2), 748--758, (2020).

\bibitem{wang2009mse}
{Wang}, Z. and {Bovik}, A.~C.
\newblock Mean squared error: Love it or leave it? a new look at signal
  fidelity measures, {\em IEEE Signal Processing Magazine\/}, {\bf 26} (1),
  98--117, (2009).

\bibitem{meirovitch2001fundamentals}
Meirovitch, L., {\em Fundamentals of Vibrations\/}, McGraw-Hill higher
  education, McGraw-Hill (2001).

\bibitem{mcintyre1986woodproperties3}
McIntyre, M.~E. and Woodhouse, J.
\newblock On measuring wood properties, part 3, {\em J. Catgut Acous. Soc.\/},
  {\bf 45}, 14--23, (1986).

\bibitem{sprossman2017materialproperties}
Sproßmann, R., Zauer, M. and Wagenführ, A.
\newblock Characterization of acoustic and mechanical properties of common
  tropical woods used in classical guitars, {\em Results in Physics\/}, {\bf
  7}, 1737 -- 1742, (2017).

\bibitem{yano1994dampingtreatment}
Yano, H., Kajita, H. and Minato, K.
\newblock Chemical treatment of wood for musical instruments, {\em The Journal
  of the Acoustical Society of America\/}, {\bf 96} (6), 3380--3391, (1994).

\bibitem{goken2018dampingmoisture}
Göken, J., Fayed, S., Schäfer, H. and Enzenauer, J.
\newblock A study on the correlation between wood moisture and the damping
  behaviour of the tonewood spruce, {\em Acta Physica Polonica A\/}, {\bf 133},
  1241--1260, (2018).

\bibitem{mihualcicua2020frequency}
Mih{\u{a}}lcic{\u{a}}, M., Stanciu, M.~D. and Vlase, S.
\newblock Frequency response evaluation of guitar bodies with different bracing
  systems, {\em Symmetry\/}, {\bf 12} (5), 795, (2020).

\bibitem{corradi2017pianotracking}
Corradi, R., Miccoli, S., Squicciarini, G. and Fazioli, P.
\newblock Modal analysis of a grand piano soundboard at successive
  manufacturing stages, {\em Applied Acoustics\/}, {\bf 125}, 113 -- 127,
  (2017).

\bibitem{schwarz1999emabasics}
Schwarz, B. and Richardson, M.~H. Experimental modal analysis, {\em Proceedings
  of CSI Reliability Week\/}, (1999).

\bibitem{chopra2019fundamentals}
Chopra, A.~K., {\em Dynamics of Structures: Theory and Applications to
  Earthquake Engineering\/}, Pearson (2006).

\bibitem{jain2005zscore}
Jain, A., Nandakumar, K. and Ross, A.
\newblock Score normalization in multimodal biometric system, {\em Pattern
  Recognition\/}, {\bf 38}, 2270--2285, (2005).

\bibitem{torres2020titian}
Torres, J.~A., Soto, C.~A. and Torres-Torres, D.
\newblock Exploring design variations of the titian stradivari violin using a
  finite element model, {\em The Journal of the Acoustical Society of
  America\/}, {\bf 148} (3), 1496--1506, (2020).

\bibitem{gruvaeus1972dendrogram}
Gruvaeus, G. and Wainer, H.
\newblock Two additions to hierarchical cluster analysis, {\em British Journal
  of Mathematical and Statistical Psychology\/}, {\bf 25} (2), 200--206,
  (1972).

\bibitem{gonzalez2020a}
Gonzalez, S., Salvi, D., Antonacci, F. and Sarti, A.
\newblock Eigenfrequency optimisation of free violin plates, {\em The Journal
  of the Acoustical Society of America\/}, {\bf 12} (1), 55--67, (2020).

\end{thebibliography}
\endgroup
\end{document}